\documentclass[a4paper,twoside]{article}

\usepackage{epsfig}
\usepackage{subcaption}
\usepackage{calc}
\usepackage{amssymb}
\usepackage{amstext}
\usepackage{amsmath}
\usepackage{amsthm}
\usepackage{multicol}
\usepackage{pslatex}
\usepackage{apalike}
\usepackage{SCITEPRESS}

\begin{document}

\title{SPIDER: Selective Plotting of Interconnected Data and Entity Relations}

\author{\authorname{Pranav Addepalli\sup{1}, Eric Wu\sup{2}, Douglas Bossart\sup{3}, Christina Lin\sup{3}, Allistar Smith\sup{4}, Ray Dos Santos\sup{5}, and Harland Yu\sup{5}}
\affiliation{\sup{1}Loudoun Academy of Science, Leesburg, VA}
\affiliation{\sup{2}University of Maryland, College Park, MD}
\affiliation{\sup{3}Virginia Tech, Blacksburg, VA}
\affiliation{\sup{4}George Mason University, Fairfax, VA}
\affiliation{\sup{5}U.S. Army Corps of Engineers – ERDC}
\email{paddepal@andrew.cmu.edu, wueric@terpmail.umd.edu, \{bossartd, christie\}@vt.edu, asmit48@masonlive.gmu.edu, \{raimundo.f.dossantos, harland.yu\}@usace.army.mil}
}

\keywords{Intelligence Analysis, Natural Language Processing, Relation Extraction, Concept Maps, Co-referencing}

\abstract{Intelligence analysts have long struggled with an abundance of data that must be investigated on a daily basis.
In the U.S. Army, this activity involves reconciling information from various sources, a process that has been automated to a certain extent, but which remains highly manual. To promote automation, a semantic analysis prototype was designed to aid in the intelligence analysis process. This tool, called Selective Plotting of Interconnected Data and Entity Relations (SPIDER), extracts entities and their relationships from text in order to streamline investigations. SPIDER is a web application that can be remotely-accessed via a web browser, and has three major components: (1) a Java API that reads documents, extracts entities and relationships using Stanford CoreNLP, (2) a Neo4j graph database that stores entities, relationships, and properties; (3) a JavaScript-based SigmaJS visualization tool for displaying the graph on the browser. SPIDER can scale document analysis to thousands of files for quick visualization, making the intelligence analysis process more efficient, and allowing military leadership quicker insights into a vast array of potentially-hidden knowledge.}

\onecolumn \maketitle \normalsize \setcounter{footnote}{0} \vfill

\section{\uppercase{Introduction}}
\label{sec:introduction}

\noindent Analysts are often tasked with fusing disparate data streams in multiple formats in an attempt to build meaningful threads of information. This process, which we call \textit{Storytelling}, is the practice of linking entities through relationships to uncover supporting facts for decision-making. In the Army, Storytelling has become increasingly significant, for example, as in the case of the Brigade S-2 staff section, whose actionable intelligence is critical to a mission’s success. A major challenge of Storytelling, however, is that analysts are often required to read, understand, and draw conclusions from hundreds of thousands of documents, a complicated and arduous manual task. Kang and Stasko (2011) found that we can help intelligence analysts by developing sophisticated analytic tools to assist and externalize their thinking process, building upon conceptual models.

In this paper, we propose a system that is capable of autonomously building a concept graph from a multitude of documents, and provides a browser-based visualization interface with a number of tools for exploratory analysis. Figure 1 shows an example of a concept map (i.e., a small graph) generated by our system. It depicts a set entities (green and black nodes) connected by edges that denote relationships. Our system is designed as a general framework of data analysis applicable to many real-world domains, which researchers can build upon for their own knowledge discovery.

\begin{figure}[!h]
  \centering
   {\epsfig{file = 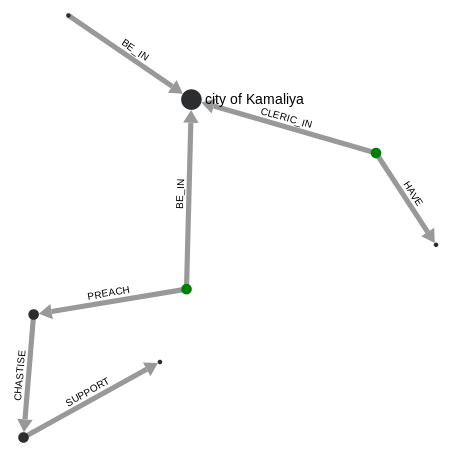, width = 7 cm}}
  \caption{An example of a Concept Map produced by our system.}
  \label{fig:ConceptMapExample}
 \end{figure}

\subsection{Limitations of Current Tools}
Existing tools for data storytelling are limited in their ability to identify relevant information to analysts. Hossain et al. \cite{Hossain12} finds that current tools, e.g., Entity Workspace \cite{Bier06}, Palantir \cite{Palantir09}, Jigsaw \cite{Stasko08}, and NetLens \cite{NetLens07}, "use entity recognizers to infer a graph of relationships between entities...In all cases, manual exploration becomes difficult as networks get dense with increasing numbers of entities." Entity Workspace lacks graph exploration capabilities as it is primarily a visualization and representation tool. Jigsaw and Netlens provide only basic exploration techniques such as traversing documents based on entity appearance and adjusting visualizations with centrality algorithms. Despite these important features, the current tools are simply unemployable for analysts. Significant shortcomings of the current tools include:
\begin{enumerate}
    \item {\em Lack of support for semantic analysis:} A substantial limitation of current tools is the absence of semantic meaning. The above tools employ entity network models where edges represent a boolean relationship between entities, i.e. the presence or absence of entities in the same sentence or document(s). They still require analysts to read documents and perform their own manual analyses. For example, the sentence "An outbreak of cholera has spread along the Dyala river in eastern Baghdad" has three entities: 'outbreak,' 'Dyala river,' and 'eastern Baghdad.' Although existing tools would connect them since they are in the same sentence, the tools would not specify the true relationships binding them, forcing analysts to read the source documents. The extracted relationships between entities should be explained by the tool in a smarter way than simple document or sentence co-occurrence criteria. Thus the need to inject semantic support into the system. \newline
    
    \item {\em Lack of support for entity disambiguation:} Current tools also fail to identify entities when referred to using pronominal references. For example, in the report "The group of soldiers left the bunker yesterday. They returned this morning," it is clear that 'the group of soldiers' and 'they' both refer to the same entity. Current tools would yield separate entities with different relationships, leaving the disambiguation task to the analyst. The lack of such capability decreases usefulness of Storytelling because of an increase in redundancy.  \newline

    \item {\em Lack of support for scaling:} The above tools fail to include the most current research from the field of data storytelling. Rather, the products are based on the features and algorithms established before the development of the product. Additionally, deploying additional features is either difficult or not possible with the above tools. There is a need for an application framework that can easily be deployed, scaled, and improved upon with current research. \newline
\end{enumerate}

\subsection{Contributions}
The above shortcomings motivated our development of SPIDER, a novel system designed to address the above issues while allowing analysts to harness the power of data storytelling. At a high level, a collection of documents is ingested, natural language processing is employed to extract entities and semantic relationships, a graph database is populated, and a web client provides graph exploration. Our contributions are:
\begin{enumerate}
    \item {\em Relation extraction} We introduce a new feature in data storytelling that defines contextual edges between entities, different from the boolean edge models employed by current tools. Rather than defining edges based on document syntax, SPIDER uses semantic analysis to define edges with specific types, which represent the contextual relationship between the entities. SPIDER's relation extraction feature explains relationships and reduces the need for analysts to refer back to the documents. With traditional tools, for example, the sentence "The men spoke to their leader" would yield entities 'the men' and 'leader' connected with a non-descriptive edge. SPIDER, on the other hand, would extract the same entities, but relate them with an edge specified as 'spoke to.' This component addresses issue (1).
    \item {\em Coreference Resolution} SPIDER employs coreference resolution to disambiguate pronominal references that represent the same entity. As a result, co-reference resolution combines entities by merging duplicates entities into a single instance. This component addresses issue (2).
    \item {\em Web Application Framework} SPIDER runs on a server as a deployable web application and allows high data loads, rather than performing computationally-costly operations on a local device. All operations except for visualization occur on the server-side. Additionally, the SPIDER system architecture allows for researchers to efficiently add new features to the application framework, especially from current research in data storytelling. This component addresses issue (3).

\end{enumerate}

\section{\uppercase{Software and Methodology}}

\noindent SPIDER is a Java web application built on the Spring Framework and running on an Apache Tomcat server. Figure 1 shows SPIDER's overall architecture which is constituted of three main components:

\noindent {\bf The Tomcat Server/Java API.} Communication between the components is handled by the Tomcat Server/Java API, which sends requests with the HTTP Protocol and parses JSON responses. The Tomcat Server/Java API also handles the Spring application's web controller, natural language processing toolkit, and graph exploration tools, which are described below. 

\noindent {\bf The Web Client} provides an upload capability for document, displays a graph of the extracted entities and relationships, and allows the user to explore the graph. 

\noindent {\bf The Neo4j Graph Database} stores the extracted entities and relationships as nodes and edges with custom properties, such as location.

\subsection{Natural Language Processing}

Stanford CoreNLP \cite{Manning14} is one of the most powerful, Java-based, linguistic analysis toolkits with support for many human languages. It was selected for this project because of its simplicity of use, robust annotators, and extensibility. The main annotators used are the named entity recognizer (NER), the coreference resolution system, and the open information extraction tool (OpenIE). 

\subsubsection{Named Entity Recognition} Stanford NER includes pre-trained models for entity recognition, one of which includes 3 classes (person, organization, and location). It is also known as CRFClassifier, as it implements linear chain Conditional Random Field (CRF) sequence models. 

\subsubsection{Coreference Resolution} The coreference annotator associates pronominal entity mentions with their nominal equivalents. Of the three algorithms - deterministic, statistical, and neural, neural has shown to be the most accurate, and thus chosen despite its longer runtime. This information is critical to SPIDER as pronominal information is of little use outside of their document or sentence contexts.

\subsubsection{Open Information Extraction} Stanford OpenIE extracts relation triples (subject, relation, object) from the text, using the results from coreference resolution to eliminate pronouns. These triples provide the raw information that is then stored, processed, and displayed by SPIDER. 

\subsection{Dominating Decision Rules}

Because of the redundancy of the information found in the relation triples generated by OpenIE, decision rules were implemented in order to reduce the noise found in the graph. The basis of developing these rules is the concept of domination. This is realized by looking at a set of relation triples that contain similar information, exploring what structural parts of these relation triples are alike and what parts are different, and deciding the best ways to eliminate weaker relation triples that will not provide as much information. The most effective rules that were incorporated into our application are as follows:

\noindent {\bf Subject-Relation Match.} Checks for relation triples containing both identical subjects and relations, and deletes the triple with the least information in the object.

\noindent {\bf Subject-Object Match.} Checks for relation triples containing both identical subjects and objects, and delete the triple with the least information in the relation.

\noindent {\bf Relation-Object  Match.} Checks for relation triples containing both identical relation and subjects, and deletes the triple with the least information in the subject.

\noindent {\bf Relation-Object Crossover.} If a triple has a relation that contains the text of another relation triple's object, delete the latter triple.

\subsection{Graph Database Platform}

\begin{figure*}[!h]
  \centering
   {\epsfig{file = 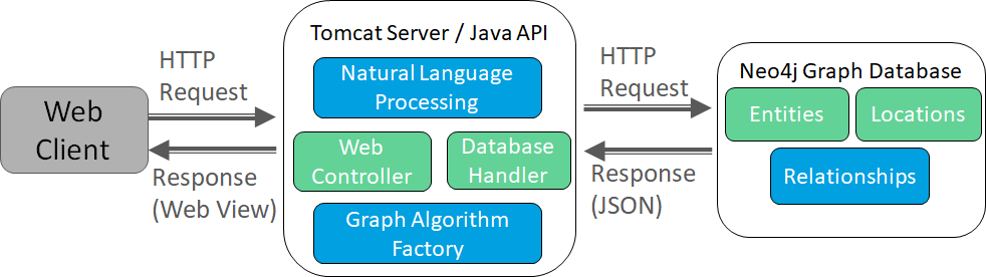, width = 15 cm}}
  \caption{SPIDER's overall architecture.}
  \label{fig:OverallArchitectureDetailed}
 \end{figure*}

Following the extraction of entities and their semantic relationships, the Neo4j graph database platform is used to store the data as a graph such that entities are represented by nodes and relationships are represented by edges. Both nodes and edges have custom properties that are assigned throughout the extraction process. Nodes are named according to the textual content from a document (e.g., "Joe"), which is used as the label during graph visualization. During the extraction process, nodes are also classified as "person" or "organization". Edges have a property of the location of the relationship, which is determined by comparing the location of the source node (the subject) and that of the target node (the object).

Connecting to the Neo4j server requires the use of the Bolt protocol operating over a WebSocket, specifically through the use of the Java Bolt driver. As the overall system architecture (Figure \ref{fig:OverallArchitectureDetailed}) shows, the Tomcat Server/Java API contains a Database Handler element. The Database Handler implements the Java Bolt driver to connect to the Neo4j graph database, send requests in the Cypher Query Language, and parse the JSON responses for use in the rest of the application.

\subsection{Graph Visualization}

SigmaJS provides the primary platform for the graph drawing element of the application. It was chosen for this project since it has a Neo4j plugin that can directly communicate with the database. Its focus on extensibility allows for more customization surrounding event bindings. This allows for node selection, manipulation, and emphasis. The Neo4j plugin for SigmaJS takes the nodes and edges created into Neo4j and produce them with sigma. 

\subsubsection{JavaScript library} The code that creates and visualizes our graph is in JavaScript, which is where we use the SigmaJS library. This is where we create the graph container and manipulate the visualization capabilities to output the sigma graph. We can change certain attributes of our graph by manipulating settings like gravity (nodes congregate in a circle or spread apart from each other), edge thickness (relationships created as clear lines) and use plugins like Force Atlas to affect the continuous graph layout and movements. This is also where we can manipulate nodes by clicking or hovering over. Similar functions are available for the edges. SigmaJS allows customization of the graph to make it visually easy to follow.

\subsection{Graph Analytics}

SPIDER provides multiple graph algorithms as well as a framework for additional functions in order to improve upon the graph exploration capability. SPIDER implements algorithms in Cypher (a NEO4J query language) and Java, and can be modified to access all algorithms already built into the Neo4j platform. These implementations are contained in the Graph Algorithm Factory element of the Tomcat Server/Java API, as seen in the overall system architecture (Figure \ref{fig:OverallArchitectureDetailed}). SPIDER currently has three primary features for graph analytics:

\subsubsection{Pathfinding} SPIDER implements Dijkstra's Shortest Path First algorithm using the Neo4j APOC plugin to find the shortest path between two nodes. A user can select two nodes and click on a button labeled “Find path” for SPIDER to calculate the shortest path and highlight it on the graph visualization.

\subsubsection{Centrality} SPIDER uses a centrality algorithm, specifically the Closeness Centrality algorithm included in the Neo4j APOC plugin, to rank nodes based on their inbound and outbound edges. A user can click a button labeled “Centrality” for SPIDER to rank all nodes and adjust node sizes on the graph visualization.

\subsubsection{Querying} SPIDER features a querying algorithm for users to be able to see only specific relationship types. For example, the user can submit a relationship of type "travel" for SPIDER to display only those edges and their corresponding source and target nodes.

\begin{figure}[!h]
  \centering
   {\epsfig{file = 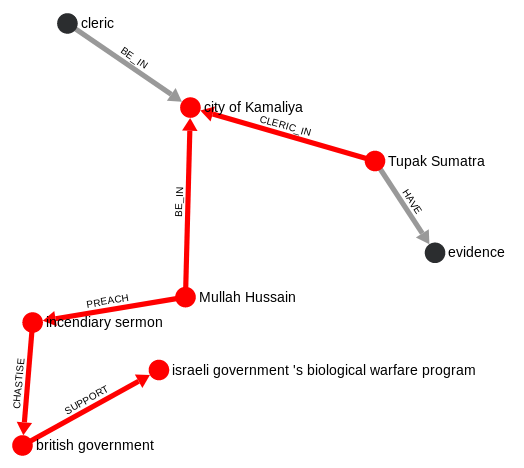, width = 5.5cm}}
  \caption{An example of pathfinding with SPIDER. "Tupak Sumatra" and "Israeli government's biological warfare program" were the two selected nodes.}
  \label{fig:Pathfinding}
 \end{figure}

\begin{figure}[!h]
  \centering
   {\epsfig{file = 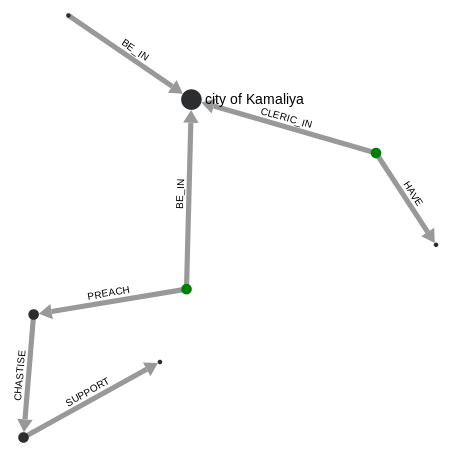, width = 5.5cm}}
  \caption{An example of the centrality algorithm with SPIDER.}
  \label{fig:Centrality}
 \end{figure}

\begin{figure}[!h]
  \centering
   {\epsfig{file = 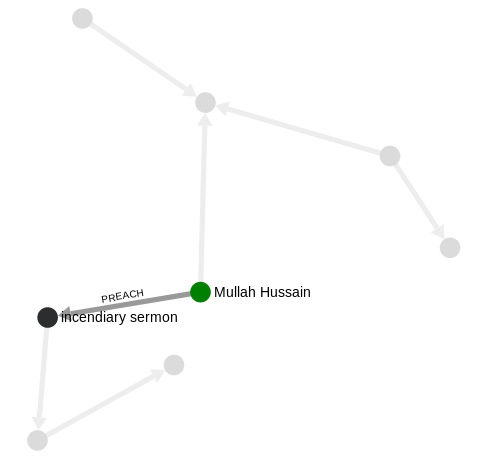, width = 5.5cm}}
  \caption{An example of querying with SPIDER. The relationship type "preach" was searched for.}
  \label{fig:Querying}
 \end{figure}

Figures \ref{fig:Pathfinding}, \ref{fig:Centrality}, and \ref{fig:Querying} show the results of the above features in SPIDER.

\section{\uppercase{Conclusions}}

We have presented a novel system for data storytelling that transforms large document collections into graphs. SPIDER is an effective tool for semantic analysis potentially useful in a wide range of applications such as Intelligence Analysis.

SPIDER was evaluated and tested with the Synthetic Counterinsurgency (SYNCOIN) dataset \cite{Graham11} and the Refined Ali Baba Group dataset \cite{Mittrick12}, resulting in a total of over 1000 military reports inspired by a four month (January 2010 to May 2010) counterinsurgency scenario centered in Baghdad, Iraq. The information is organized into 8 threads (Bio-Weapons, Baath'ist Resurgent, Iranian Special Group, Sectarian Conflict, Sunni Criminal, Rashid IED, Ali Baba Group, and Ali Baba Ground Truth). SPIDER was able to effectively extract and visualize information from the entire dataset with an approximate extraction efficiency of 75\%.

SPIDER's contribution to the field of intelligence analysis and visualization includes the ability to extract semantic relationships between entities from text documents rather than simply using boolean relationships between entities. This reduces the need for manual searches for relationships between entities, as well as provides more complete relationships for constructing information threads. SPIDER also features entity disambiguation via coreference resolution, allowing for the construction of information threads across multiple documents and reports without redundancy. SPIDER's implementation of Decision Rule theory, specifically the concept of domination, improves upon the current capabilities of natural language processing software by significantly reducing the number of extracted relation triples while maintaining the accuracy of the information. Specifically, SPIDER's implementation of decision rules narrowed down the Stanford CoreNLP extracted triples from the full SYNCOIN dataset from 8,603 to 4,209 relation triples, approximately 51.1\%, with potential to improve this reduction even further. Ultimately, as information accuracy and clarity are the most important features of an application for intelligence analysts, SPIDER's use of semantic relationship extraction, coreference resolution, and dominating decision rules is crucial for the future of intelligence analysis tools.

Additionally, SPIDER's overall system architecture provides a platform for researchers in data storytelling to easily build upon and integrate new features. Its simple characteristic of being deployable as a web application increases the tool's accessibility , and the nature of the architecture reduces the need for computationally powerful devices for analysis. SPIDER's ability to ingest and extract large document collections aids in simplifying the manual analysis process, and its visualization platform with graph analytics and exploration tools substantially expands the potential of intelligence analysts for fields such as quantitative threat prediction, anomaly detection, document visualization, and graph theory.

Future work includes improving the UI of the application, training the natural language processing model on a military-style dataset, geotagging entities with real-world spatial and temporal coordinates, ranking entities to produce advanced stories with similarity metrics using algorithms like ConceptRank \cite{DosSantos16}, and adding additional graph traversal and analytics algorithms such as link prediction to improve the analysis capabilities of SPIDER.

\section*{\uppercase{Acknowledgements}}

We would like to thank the employees of the U.S. Army Corps of Engineers -- Geospatial Research Laboratory, especially our mentors, Dr. Raimundo F. Dos Santos and Harland Yu, for their continued support throughout the duration of our project.

\bibliographystyle{apalike}
{\small
\bibliography{SPIDER}}

\end{document}